\documentclass[twocolumn,showpacs,preprintnumbers,amsmath,amssymb]{revtex4}
\usepackage{amsmath}
\usepackage{amsfonts}
\usepackage{amssymb, multirow}
\usepackage{pdfsync}
\usepackage{epsfig}
\usepackage{graphicx}
\usepackage{subfigure}
\usepackage{nicefrac}

\newcommand{\be}{\begin{equation}}\newcommand{\ee}{\end{equation}}
\newcommand{\bea}{\begin{eqnarray}} \newcommand{\eea}{\end{eqnarray}}
\newcommand{\ba}[1]{\begin{array}{#1}} \newcommand{\ea}{\end{array}}

\long\def\symbolfootnote[#1]#2{\begingroup%
\def\thefootnote{\fnsymbol{footnote}}\footnote[#1]{#2}\endgroup} 

\def\wdg{{\scriptstyle \wedge}}

\newcommand{\cN}{{\cal N}}

\newcommand{\ft}[2]{{\textstyle\frac{#1}{#2}}}

\def\bfone{\relax{\rm 1\kern-.35em 1}}

\begin{document}

%\preprint{}

\title{All $G_2$ invariant critical points of maximal supergravity}
% Force line breaks with \\

\author{Andrea Borghese, Adolfo Guarino and Diederik Roest}
% \altaffiliation[Also at ]{Physics Department, XYZ University.}
%Lines break automatically or can be forced with \\
%\author{Second Author}%
% \email{Second.Author@institution.edu}
\affiliation{%
~ \\
Centre for Theoretical Physics, University of Groningen, Nijenborgh 4 9747 AG Groningen, The Netherlands
%\smallskip
%This line break forced with \textbackslash\textbackslash
}%

\begin{abstract}

We perform an exhaustive classification of $G_2$ invariant extrema of the most general gauged $\cN = 8$ supergravity in four dimensions. They comprise four branches of Anti-de Sitter solutions labelled by a single parameter. Interestingly, while the gauge groups vary with the parameters, the mass spectra are invariant. One of these is a new non-supersymmetric yet stable point. Our analysis includes the recently proposed family of $SO(8)$ gauged supergravities and more.

%An article usually includes an abstract, a concise summary of the work
%covered at length in the main body of the article. It is used for
%secondary publications and for information retrieval purposes. Valid
%PACS numbers may be entered using the \verb+\pacs{#1}+ command.
\end{abstract}

\pacs{04.65.+e, 11.25.Tq}
% PACS, the Physics and Astronomy Classification Scheme.
%\keywords{Suggested keywords}%Use showkeys class option if keyword
                              %display desired
\maketitle

\section{Introduction}

Maximal gauged supergravity in four dimensions \cite{N=8a} has played a pivotal role in many of the developments in string and M-theory. In particular, its $SO(8)$ incarnation arises as an $S^7$ compactification of 11D supergravity \cite{deWit1, deWit2}. Moreover, it provides the gravity dual to a family of 3D conformal field theories \cite{Bagger:2007jr, Aharony}. Some of the critical points of the theory have also been employed in the AdS/CMT correspondence \cite{Bobev:2011rv}. Finally, changing the gauge group to non-compact versions such as $SO(4,4)$ or $SO(5,3)$ also allows for De Sitter solutions \cite{HW84}.

It is remarkable that, after a few decades of intense research, the theory still offers surprises. New critical points of the $SO(8)$ theory were found with numerical methods \cite{Fischbacher}. Moreover, contrary to expectation, it was realised recently that one of the known non-supersymmetric critical points was actually perturbatively stable \cite{stable}. Yet much more recently, it was found that there is in fact a one-parameter family of $SO(8)$-gauged supergravities \cite{Inverso}.

In view of these developments, it would be advantageous to have an overview of {\it all} critical points of {\it all} gauged supergravities. For the standard $SO(8)$ theory, all critical points preserving e.g.~an $SU(3)$ subgroup of $SO(8)$ have been classified \cite{Warner1983}. In this letter we will extend this result by classifying all critical points preserving a $G_2$ subgroup of all gauge groups of maximal supergravity. This includes both the standard $SO(8)$ plus its one-parameter generalisation of \cite{Inverso}, but also allows for other gauge groups, as we will see.

Our approach, as proposed and applied in an $\cN = 4$ context in \cite{DGR} and subsequently in an $\cN = 8$ context in \cite{DAI}, will be crucially different from the usual search for critical points. Normally one chooses the gauge group and structure constants, calculates the corresponding scalar potential and scans the moduli space for critical points of that potential. Instead, we will choose the critical point to be at the origin. This does not entail a loss of generality in the case of a homogeneous scalar manifold, such as maximal supergravity has. Subsequently we will scan for all structure constants that are consistent with having a critical point at the origin. This allows us to calculate the gauge group and mass spectra of this point. The advantage of this approach is a massive reduction of calculational complexity. While solving field equations in general quickly involves higher-order equations, in our approach one encounters only up to quadratic equations. A disadvantage is that it can be difficult to see which critical points belong to the moduli space of the same theory. This requires the structure constants of the two solutions to be related via a duality transformation, which in general is a non-trivial analysis.

Specialising to $D=4$ maximal supergravity, the scalar manifold is $E_{7(7)} / SU(8)$ and the vectors span the $\bf 56$ irrep of $E_{7(7)}$. The most general gaugings are parametrised by the so-called embedding tensor \cite{ET1,ET2}, which are a duality covariant generalisation of the structure constants. For maximal supergravity these span the $\bf 912$ of $E_{7(7)}$ \footnote{By gauging the trombone symmetry one can introduce an additional $\bf 56$ \cite{LeDiffon}, which our method can incorporate as well. It will not play any role in what follows as it does not have any $G_2$ singlets.}. For consistency of the gauging one needs to impose the duality covariant Jacobi identities, the so-called quadratic constraints (QC), living in the $\bf 133$ and the $\bf 8645$.

The restriction to the origin of moduli space forces one to give up $E_{7(7)}$ covariance and instead employ the maximal compact subgroup $SU(8)$. The embedding tensor decomposes into two complex irreps, $\bf 36$ and $\bf 420$, which will be denoted by $A^{IJ}$ and $A_I{}^{JKL}$, where $I=1,\ldots,8$. These are subject to the decomposition of QC into $SU(8)$, whose explicit form can be found in \cite{4DMS}. Finally, in order for the origin to be a critical point, one needs to impose the equations of motion (EOM) of the scalars. These live in the ${\bf 70}_+$ irrep and are given by 
 \begin{align}
 & A^{R}{}_{[IJK} A_{L]R} + \tfrac{3}{4} A^{R}{}_{S[IJ} A^{S}{}_{KL]R} = \label{Field equations} \\
 & = - \tfrac{1}{4!} \epsilon_{IJKLMNPQ} \left( A_{R}{}^{MNP} A^{QR} + \tfrac{3}{4} A_{R}{}^{SMN} A_{S}{}^{PQR} \right) \, . \notag
 \end{align}
It can be seen that the combined QC and EOM have the following two discrete symmetries:
 \begin{align}
  (A^{IJ}, A_I{}^{JKL}) & \rightarrow (i A^{IJ}, -i A_I{}^{JKL}) \,, \notag \\
  &  \rightarrow (A^{IJ}, A_I{}^{JKL})^* \,.  \label{discrete}
 \end{align}
The latter are a natural consequence of the fact that the QC and EOM are real irreps of $E_{7(7)}$ and $SU(8)$, respectively, and hence cannot distinguish between e.g.~the ${\bf 36}$ and the $\overline{\bf 36}$. 

Given a solution to the combined system of QC and EOM, the scalar mass spectrum can be calculated as eigenfunctions of the Hermitian matrix
\begin{align}
& m^{2}_{IJKL}{}^{MNPQ} = \delta_{IJKL}^{MNPQ} \left( \tfrac{1}{6} \, |A_{2}|^{2} - 3 \, |A_{1}|^{2} \right) + \notag \\ 
& + 20 \, \delta_{[IJK}^{[MNP} A_{L]R} A^{Q]R} + 6 \, \delta_{[IJ}^{[MN} A_{K}{}^{RS|P} A^{Q]}{}_{L]RS} + \notag \\
& - \tfrac{2}{3} \, \delta_{[I}^{[M} A_{R}{}^{NPQ]} A^{R}{}_{JKL]} - \tfrac{2}{3} \, A_{[I}{}^{[MNP} A^{Q]}{}_{JKL]} \,.
\end{align}
Similarly, the mass spectrum for the vectors is determined by \cite{LeDiffon2}
\begin{align}
m^2 =
\left(
\begin{array}{cc}
m^2_{IJ}{}^{KL} & m^2_{IJKL}
\\
 {m^2}^{IJKL}&
 {m^2}^{IJ}{}_{KL}
\end{array}
\right)
\,,
\label{M_vector}
\end{align}
with
\begin{align}
m^2_{IJ}{}^{KL}
= &
-\ft16 A_{[I}{}^{NPQ} \delta_{J]}^{[K} A^{L]}{}_{NPQ}
+\ft12 A_{[I}{}^{PQ[K}A^{L]}{}_{J]PQ} \,, \nonumber\\
m^2_{IJKL} =& \,
\ft1{36}  A_{[I}{}^{PQR} \epsilon_{J]PQRMNS[K} A_{L]}{}^{MNS} \,,
\end{align}
whose eigenvalues consist of 28 masses for the physical gauge vectors and 28 zeroes for the non-physical dual vectors.
Finally, supersymmetry requires one or several of the eigenvalues of $A^{IJ}$ to coincide (up to a phase) with $\sqrt{-V/6}$, where $V = -\tfrac34 |A_1|^2 + \tfrac{1}{24} |A_2|^2$ is the value of the scalar potential in the critical point. 

\section{$G_2$ invariant classification}

Any critical point that preserves a $G_2$ subgroup of any gauge group has an embedding tensor that, after taking the scalar dependence into account, is $G_2$ invariant. Assuming this critical point to be at the origin of moduli space (which  is fully general, as we explained in the introduction) implies that the embedding tensor itself must be $G_2$ invariant. The search for all critical points with such invariance thus translates into the search for all embedding tensors that are $G_2$ invariant. In order for these to correspond to both a consistent gauging and a critical point, one has to impose the QC and the EOM. For this reason our classification only involves quadratic expressions of the parameters that one is solving for.

In order to parametrise $G_2$ invariant tensors, we split indices according to
 \begin{align}
   I = ( 1, m) \,, 
 \end{align}
where $I,J,...$ is the fundamental of $SU(8)$ and $m,n,...$ is the fundamental of $G_2$. The latter is also the fundamental of $SO(7)$ when embedded in $SO(8)$ in the standard way, in which the 8D vector decomposes in a 7D scalar and vector.
Then one can define the following subgroups. $G_2$ is the subgroup of $SO(7)$ that leaves the following three-form and its dual four-form invariant: 
\begin{align}
\varphi = & e^{2} {\wdg} e^{3} {\wdg} e^{4} + e^{2} {\wdg} e^{5} {\wdg} e^{8} + e^{2} {\wdg} e^{7} {\wdg} e^{6} +\notag \\
  & + e^{3} {\wdg} e^{5} {\wdg} e^{7} + e^{3} {\wdg} e^{6} {\wdg} e^{8} + e^{4} {\wdg} e^{6} {\wdg} e^{5} + e^{4} {\wdg} e^{7} {\wdg} e^{8} \notag \, . \\
  * \varphi = & e^{5} {\wdg} e^{6} {\wdg} e^{7} {\wdg} e^{8} + e^{3} {\wdg} e^{4} {\wdg} e^{6} {\wdg} e^{7} + \notag \\ 
  & + e^{3} {\wdg} e^{4} {\wdg} e^{8} {\wdg} e^{5} + e^{2} {\wdg} e^{4} {\wdg} e^{6} {\wdg} e^{8} + e^{2} {\wdg} e^{4} {\wdg} e^{5} {\wdg} e^{7} + \notag \\
  & + e^{2} {\wdg} e^{3} {\wdg} e^{8} {\wdg} e^{7} + e^{2} {\wdg} e^{3} {\wdg} e^{5} {\wdg} e^{6} \, .
\end{align}  
Secondly, due to triality one can define two other $SO(7)$ subgroups of $SO(8)$, corresponding to the one where either the positive or the negative chirality spinor decomposes into a 7D scalar and vector. These are defined by requiring the invariance of the (anti-)self-dual four-form
 \begin{align}
   e^1 {\wdg} \varphi \pm * \varphi \,, \label{ASD4}
 \end{align}
and will be denoted by $SO(7)_\pm$.

Decomposing the $\bf 36$ and $\bf 420$ of $SU(8)$ into $G_2$, one finds two and three singlets, respectively. We will parametrise these with the following Ansatz:
 \begin{align}
  & A^{11} = \alpha_1 \,, \quad 
  A^{mn} = \alpha_2 \delta^{mn} \,, \notag \\
  & A_1{}^{mnp} = \beta_1 \varphi^{mnp} \,, \quad 
  A_m{}^{1np} = \beta_2 \varphi_{m}{}^{np} \,, \notag \\
 &   A_{m}{}^{npq} = \beta_3 (* \varphi)_m{}^{npq} \,.
 \end{align}
where all indices of $\varphi$ and $* \varphi$ are raised and lowered with the $SO(7)$ metric $\delta^{mn}$ (which is also an invariant tensor of $G_2$). For special values of these parameters the invariance can be enhanced. For instance, when $\vec \alpha = (\alpha, \alpha)$ and $\vec \beta = (\beta, - \beta, \pm \beta)$, the embedding tensor can be written in terms of the $SO(8)$ invariant metric and the (anti-)self-dual four-form \eqref{ASD4}, and thus has an $SO(7)_{\pm}$ invariance. Moreover, when $\vec \alpha = (\alpha, \alpha)$ and $\vec \beta = (0, 0, 0)$, the invariance group is actually the largest possible, being $SO(8)$.

Plugging the most general Ansatz with five complex parameters into the QC and the EOM one gets a number of quadratic constraints on these parameters. As explained in more detail in \cite{DGR}, these are amenable to an exhaustive analysis by means of algebraic geometry techniques, in particular prime ideal decomposition, and the corresponding code Singular \cite{singular}. In this way we find the four branches of solutions listed below, all corresponding to Anti-de Sitter space-times. In all cases we will omit an overall scaling of the solutions and use $SU(8)$ to set the phases of $\alpha_1$ and $\alpha_2$ equal. All four branches have a single remaining parameter. They are either $\cN= 0,1$ or $8$, depending on how many of the supersymmetry conditions, which now read
 \begin{align}
  8 |\alpha_{1,2}|^2 = |\alpha_1|^2 + 7 |\alpha_2|^2 - \tfrac73 |\beta_1|^2 - 7 |\beta_2|^2 - \tfrac{28}{3} |\beta_3|^2 \,, \notag
 \end{align}
are satisfied: \\

\noindent
$\bullet$ The first branch is $\cN = 8$ and reads
 \begin{align}
\vec \alpha = (e^{i \theta}, e^{i \theta}) \,, \quad \vec \beta = (0, 0, 0) \,.
 \end{align}
All solutions are $SO(8)$ invariant and preserve $\cN = 8$. They correspond to the origin of the standard $SO(8)$ gauging and its one-parameter generalisation. As also noted in \cite{Inverso}, the scalar mass spectrum is equal for the entire branch and given by
 \begin{align}
  m^2 L^2 = - 2  \; (\times 70) \,,
 \end{align}
in terms of the AdS radius $L^2 = -3 / V$. Similarly, the vector mass spectrum reads
 \begin{align}
  m^2 L^2 = 0  \; (\times 56) \,,
 \end{align}
of which the 28 physical ones are gauge vectors of $SO(8)$. \\

\noindent
$\bullet$ 
The second branch is $\cN = 1$ and is given by
\begin{align}
 & \vec \alpha = ( - 2 \, e^{-5 i \theta}, \sqrt{6} \,  e^{- 5 i \theta}) \,, 
 \notag \\
 & \vec \beta = (0, \sqrt{2/3} \, e^{- i \theta}, e^{3 i \theta}) \,.
\end{align}
For all values of the parameter, the invariance group is $G_2$ and the mass spectrum reads
 \begin{align*}
 (4 \pm \sqrt{6} ) \; (\times 1) , \quad 0 \; (\times 14) , \quad - \tfrac{1}{6} (11 \pm \sqrt{6}) \; (\times 27) \, .  
 \end{align*}
This coincides with the $G_2$ invariant mass spectra of the standard $SO(8)$ theory. The latter corresponds to a particular value of $\theta$.  Other values include the one-parameter generalisation of \cite{Inverso} and possibly more. In this case the vector masses are given by
 \begin{align}
   0  \; (\times 42) , \quad \tfrac12 ( 3 \pm \sqrt{6}) \; (\times 7) \,.
 \end{align}
Half of the physical vectors are therefore massive, while the other half correspond to the $G_2$ gauge vectors.
 \\ 

\noindent
$\bullet$
The third branch is $\cN = 0$ and reads
 \begin{align}
  & \vec \alpha = (3 e^{-3 i \theta},  3 e^{- 3 i \theta} ) \,, \notag \\
 &   \vec \beta = (- e^{i \theta}, e^{i \theta}, \mp e^{i \theta} ) \,.
 \end{align}
The stability subgroup in this case is given by $SO(7)_\pm$. The mass spectrum is independent of the parameter and reads
 \begin{align*}
 6 \; (\times 1) , \quad 0 \; (\times 7) , \quad - \tfrac{6}{5} \; (\times 35) , \quad - \tfrac{12}{5} \; (\times 27) \, .
 \end{align*}
The lowest of the eigenvalues violates the Breitenlohner-Freedman bound $m^2 L^2 \geq - \tfrac94$ and hence this branch is perturbatively unstable. This spectrum coincides with the $SO(7)_\pm$ invariant mass spectra of the standard $SO(8)$ theory. Again, the latter corresponds to a single point in a one-dimensional parameter space of non-supersymmetric $SO(7)_\pm$ invariant critical points, as also found in \cite{DAI}. Turning to the vector masses, we find
 \begin{align}
     0  \; (\times 49) , \quad \tfrac{12}{5} \; (\times 7) \,,
 \end{align}
of which the physical ones are the $SO(7)_\pm$ gauge vectors and 7 massive ones.
\\

\noindent
$\bullet$ 
The last branch is $\cN= 0 $ as well and is given by
\begin{align}
 & \vec \alpha = (\sqrt{3} \, e^{- 3 i \theta}, - e^{ - 3 i \theta}) \,, \notag \\
 & \vec \beta = (e^{i \theta}, \tfrac13 \sqrt{3} e^{i \theta}, 0) \,.
 \end{align}
The invariance group is  $G_2$ and the mass spectrum reads
 \begin{align}
 6 \; (\times 2) , \quad 0 \; (\times 14) , \quad - 1 \; (\times 54) \, .   \label{news}
 \end{align} 
In this case all eigenvalues satisfy the Breitenlohner-Freedman bound, and hence this family of critical points is non-supersymmetric and nevertheless perturbatively stable. Previously known examples of stability without supersymmetry were isolated points with smaller symmetry groups \cite{stable, exceptional}. Finally, we find
 \begin{align}
     0  \; (\times 42) , \quad 3 \; (\times 14) \,,   \label{newv}
 \end{align}
for the vector masses in this case.

\bigskip

To our knowledge, the last case presents a new mass spectrum, which does not arise in the standard $SO(8)$ theory. However, we find that it does occur in the theory of \cite{Inverso}. In particular, from the $A^{IJ}$ tensor of that paper, we have verified that their pair of new $G_2$ invariant critical points preserve $\cN = 1$ and $\cN = 0$, respectively. The first therefore belongs to our second branch of solutions. The other one indeed has the ratio $|\alpha_1 / \alpha_2| = \sqrt{3}$ and therefore belongs to the last branch. This implies that {\it all four mass spectra surface in some incarnation of $SO(8)$ maximal supergravity}. Moreover, it provides additional evidence that the family of \cite{Inverso} describes inequivalent theories.

\section{Discussion}

We have performed an exhaustive classification of critical points that preserve at least a $G_2$ subgroup of the gauge group. As we have seen, these split up in four branches of solutions: a fully supersymmetric $SO(8)$ branch, an $\cN = 1$ supersymmetric $G_2$ branch and supersymmetry breaking $SO(7)_\pm$ and $G_2$ branches. We have argued that all branches contain minima of some maximal $SO(8)$ gauged supergravity, including the novel spectra \eqref{news} and \eqref{newv}. While all branches of solutions have a single free parameter, both the scalar and vector mass spectra do not depend on this.

Our findings answer questions while raising others. In the first class is the mass and supersymmetry of the different critical points of \cite{Inverso}. The results above unambiguously demonstrate that the mass spectra of the $G_2$ invariant critical points of the family of $SO(8)$ gauged supergravity theories are equal for all values of the parameters, and thus given by the four spectra above. Moreover, of the two new $G_2$ invariant extrema of \cite{Inverso}, one belongs to the second branch and one to the fourth.

Naturally, the above prompts the question why the spectra are this simple: can one understand their parameter independence from e.g.~symmetry principles? Similarly, it would be interesting to investigate how general this statement is: does this only hold for all $G_2$ invariant points, or in fact for a larger class and possibly all? The most general $SU(3)$ invariant critical points would be a natural stepping stone in trying to answer this question. If it were the case that the mass spectra of all critical points of the one-parameter family of $SO(8)$ gaugings are parameter independent, it would be interesting to think about other physical quantities that do depend on it, such as flows between different extrema. 

A similar point applies to our branches of solutions: as the mass spectra are insensitive to the parameters, one could wonder to what extent the parameters label different solutions. We will show that indeed there are physical changes when traversing the parameter space. To this aim we have calculated the eigenvalues of the Cartan-Killing metric, from which the full gauge group (and not only the invariance group of the critical point) can be derived. Again, this is outlined in \cite{DGR} and we employ the mapping given in \cite{exceptional}. For the $SO(8)$ invariant critical points, we find that all 28 eigenvalues are negative for all values of $\theta$, as required by the $SO(8)$ gauge group. For the $SO(7)_+$ critical points, things are more interesting. A set of 21 eigenvalues is always negative, corresponding to the preserved part of the gauge group. The remaining seven are either all negative, zero or positive, as a function of $\theta$, leading to the following gauge groups:
 \begin{alignat}{2}
   \theta & \in [0, {\rm arccos}(\sqrt{\tfrac16 (3 + \sqrt{5})}): \quad &  \mathcal G  & = SO(7,1) \,, \notag \\
    & = {\rm arccos}(\sqrt{\tfrac16 (3 + \sqrt{5})}: \quad &     & = ISO(7) \,, \notag \\
    & \in ({\rm arccos}(\sqrt{\tfrac16 (3 + \sqrt{5})}, \tfrac14 \pi]: \quad &     & = SO(8) \,. \notag 
 \end{alignat}
The gauge group therefore changes from compact to non-compact and vice versa, while passing trought an In{\"o}nu-Wigner contracted point.
Other values for $\theta$ can be related to those in the interval $[0,\tfrac14 \pi]$ via the discrete symmetries \eqref{discrete}. The situation for $SO(7)_-$ gaugings is similar: we find
 \begin{alignat}{2}
   \theta & \in [0, {\rm arccos}(\sqrt{\tfrac56}): \quad &  \mathcal G  & = SO(8) \,, \notag \\
    & = {\rm arccos}(\sqrt{\tfrac56}): \quad &     & = ISO(7) \,, \notag \\
    & \in ({\rm arccos}(\sqrt{\tfrac56}), \tfrac14 \pi]: \quad &     & = SO(7,1) \,, \notag 
 \end{alignat}
for the same interval. These results tie in with those of \cite{DAI}, where it was found that the three gauge groups mentioned above have $SO(7)_\pm$ invariant critical points with identical scalar mass spectra. For the $G_2$ invariant branches we find the same pattern of 21 negative and 7 indefinite eigenvalues as a function of $\theta$. For the supersymmetric case, this leads to
  \begin{alignat}{2}
   \theta & \in [0, 0.06 \pi): \quad &  \mathcal G  & = SO(8) \,, \notag \\
    & \simeq 0.06 \pi : \quad &     & = ISO(7) \,, \notag \\
    & \in (0.06 \pi, 0.19 \pi): \quad &     & = SO(7,1) \,, \pagebreak[4] \notag  \\
    & \simeq 0.19 \pi: \quad &     & = ISO(7) \,, \notag \\
    & \in (0.19 \pi, \tfrac14 \pi]: \quad &  & = SO(8) \,,
 \end{alignat}
while for the non-supersymmetric case, the sequence is the same but transitions take place around $0.02 \pi$ and $0.23 \pi$. Thus we find similar patterns of transitions from $SO(8)$ to $SO(7,1)$ via their common contracted version $ISO(7)$ for all four branches apart from the first.

The additional parameter of \cite{Inverso}, specifying the embedding of the $SO(8)$ gauge vectors into the $\bf 56$ of the electro-magnetic duality group, is in some respects similar to the De Roo-Wagemans angle that one can introduce in half-maximal supergravity \cite{deRoo}. A crucial difference, however, is that the latter angle can only be introduced for a semi-simple gauge group, where the angle describes the electro-magnetic ``mismatch'' between the different factors of the semi-simple gauge group. An overall phase for the entire gauge group corresponds to a $U(1)$ transformation that is contained in the R-symmetry group, and consequently does not affect physics. One could therefore expect that the additional possibility to introduce such an overall phase in maximal supergravity is related to the R-symmetry being $SU(8)$ rather than $U(8)$, and that the missing $U(1)$  exactly allows for the phase of \cite{Inverso}. 
It would be interesting to investigate what this interpretation implies for theories with less supersymmetry, which do have a proper $U(\cN)$ R-symmetry group.

We hope to come back to some of these points in the near future.

%\tableofcontents

%\smallskip

\section*{Acknowledgements}

%\noindent
%{\bf Acknowledgements:} 

We would like to thank Gianguido Dall'Agata, Giuseppe Dibitetto, Gianluca Inverso, Henning Samtleben and Mario Trigiante for  useful discussions, and Oscar Varela for a very interesting collaboration at the early stages of this project. This research is supported by a VIDI grant from NWO.

%\providecommand{\href}[2]{#2}\begingroup\raggedright\begin{thebibliography}{10}

%\endgroup

\end{document}